\documentclass[preprintnumbers,article,amsmath,amssymb,floatfix,10pt,prd,onecolumn,
superscriptaddress,nofootinbib]{revtex4}
\usepackage{bbm}
\usepackage{amsfonts}
\usepackage{mathrsfs}
\usepackage{latexsym}
\usepackage[cp1251]{inputenc}
\usepackage{epsfig}
\usepackage{epstopdf}
\usepackage{graphicx}
\usepackage{amssymb}
\usepackage{amsmath}
\usepackage{dcolumn}
\usepackage{bm}
\usepackage{color}
\usepackage{comment}
\usepackage{xcolor}

\begin{document}
\title{\bf Charged Anisotropic Finch-Skea-Bardeen Spheres }

\author{M. Farasat Shamir}
\email{farasat.shamir@nu.edu.pk}\affiliation{National University of Computer and
Emerging Sciences, Islamabad,\\ Lahore Campus, Pakistan.}
\author{G. Mustafa}
\email{gmustafa3828@gmail.com}\affiliation{Department of Mathematics, Shanghai University,
Shanghai, 200444, Shanghai, People's Republic of China}
\author{Mushtaq Ahmad}
\email{mushtaq.sial@nu.edu.pk}\affiliation{National University of Computer and
Emerging Sciences,\\ CFD Campus, Pakistan.}

\begin{abstract}
This manuscript explores the compact geometries by employing Karmarkar condition with the charged anisotropic source of matter distribution. For this purpose, we consider an explicit model by indulging $\mathrm{g}_{rr}$ metric potential obeying the Karmarkar condition. Moreover, we ansatz the time metric co-efficient following the approach by Adler \cite{Adler}.
The crucial aspect of present investigation is the implication of the Bardeen model as an outer spacetime \cite{Bardeen}. Implementation of Bardeen approach turns out to be very interesting as this corresponds to the magnetic mono-pole gravitational remnants emerging from some particular non-linear electrodynamics. Detailed analysis supported by their corresponding plots of the profiles of the pressure profiles, energy density, charged density, anisotropy function, electric field attributes, energy bounds, redshift function, compactness parameter, stability and adiabatic index has been provided. It is important to mention here that our obtained solutions are physically viable and are well stable.
\\

\end{abstract}

\maketitle

\date{\today}

\section{Introduction}
\label{intro}

Compact astrophysical objects, such as black holes, neutron stars, quark stars, and gravastars have been the center of attention of the researchers to investigate some specific properties of gravitational fields. The phenomenon of the gravitational collapsing has been studied largely after the advent of the theory of general relativity, however, we are still ignorant of the crucial facts depicting the true nature of the compact stars. One thing which describes the compact stars is their massive nature and small radii due to the huge density. In literature, all these objects, except the black holes, are known to be  the degenerate objects . Schwarzschild in 1916 started an era of studying relativistic stellar models by exploring the only exact solution of the Einstein’s field equation. In modern day cosmology, exact stellar geometries admitting a diversity of the physical constraints are vitally important in structuring the contemporary gravitational physics. In particular, the study of compact stars solutions by employing different physical conditions remains an attractive matter of discussions for the last few decades. Some important literature on the subject is given below.

Spherically symmetric equilibrium structures for the polytropic matter content non-minimally amalgamated to an exterior chameleon scalar field have been studied by Folomeev et al. \cite{1a}. They also discussed the stability criteria and concluded that the stellar structures had statically regularized asymptotically flat geometry. Dzhunushaliev et al. \cite{2a} used scalar field for gravitating symmetric configurations which conceded the simple stability trials. Jetzer \cite{3a} explored the cold stellar structures developed from bosons and the fermions and discussed their dynamical stability. In some other work, Jetzer and collaborators \cite{4a} investigated dynamical instability of the statically realistic scalar field by incorporating the Einstein-Klein-Gordon geometry. Spherically symmetric stellar remnants have been explored by implementing some standard models to obtain some exact solutions and the corresponding outcomes exhibited that energy density, radial and transversal pressure attributes persisted finitely positive \cite{MaK}. Some important features of anisotropic structures along the cosmological constant have been conveyed by Hossein et al. \cite{Hossein}. Some interesting outcomes with reference to the geometries of slowly spinning neutron stars are revealed with the assumption of diverse equation of state parameters \cite{32e}. Some more notable aspects of compact structures are stated with some imperative $f(R,\mathscr{G})$ gravity models \cite{14a}. It is revealed that explicit $f(R,\mathscr{G})$ gravity models involving charge may deliver some cosmological geometries that admit the observational statistis\cite{14b}.
Similarly, many works witnessed the use of Karmarkar geometry \cite{76} to investigate the cosmological solutions and nature of stellar objects. In a recent paper \cite{14c}, we have studied the compact stars in connection to the observational data (radii and mass) by engaging the Karmarkar condition under  $f(\mathscr{G},\mathcal{T})$ gravity background and it is conferred that the attained solutions are physically feasible with well-behaved nature. For a better understanding, some important papers can be studied in detail \cite{Karmarkar1}-\cite{Karmarkar6}.

Usually, the constitution of a black hole comprises a horizon possessing a singularity where the scalar curvature becomes infinite. Nevertheless, the presence of the black holes without having any singularity at the core has also been reported which are referred to as the ``regular" black holes. Bardeen was the pioneer who attained a black hole geometry without possessing any singularity. Such geometrical structures in the literature are acknowledged as the Bardeen black hole \cite{Bardeen}. Moreno and his collaborator Sarbach \cite{Moreno} explored the stability attributes of the Bardeen black hole structures and other similar structures. The geodesic structures of test particles concerning the Bardeen geometry were analyzed by Zhou and his co-authors \cite{Zhou}. In a recent paper \cite{we}, we proposed a novel class of charged compact objects by incorporating the conformal Killing symmetries through the Bardeen model to exhibit exterior metric comparison.

Enthused by the fascinating character of the Bardeen geometry, we invest our efforts to investigate the compact stellar geometries by introducing the Bardeen geometry as an outer spacetime. This will make the analysis interesting as the employment of the Bardeen approach proves to be very fascinating as this is related to the magnetic monopole gravitational remnants evolving from some certain non-linear electrodynamics. In particular, we consider charged anisotropic source of fluid for the present study. The paper is organized as follows: In Sec \textbf{II}, we provide basic mathematics of field equations with the Karmarkar geometry to evaluate the profile of metric potential. Moreover, physical quantities like energy density, pressure components, and expressions for charge density are discussed. Sec \textbf{III} offers a discussion of the Bardeen model as comparitive exterior geometry with relative conditions. Sec \textbf{IV} is devoted to present some physical attributes of the relativistic. Summary and concluding comments is comprised in the last Sec. As far we know, this work to implement the Bardeen geometry in investigating the charged anisotropic spheres satisfying the Karmarkar condition.

\section{Basic Field Equations}

The following spacetime describes the geometry for our ongoing work

\begin{equation}\label{1}
ds^{2}=e^{\lambda(r)}dr^2+r^{2}d\theta^{2}+r^2sin^{2}\theta d\phi^{2}-e^{\nu(r)}dt^{2}.
\end{equation}
The anisotropic matter content distribution represented by the charged stress-energy tensor is provided as
\begin{equation}\label{4002}
\mathcal{T}_{\chi\gamma}=(\rho+p_{t})\upsilon_{\chi}\upsilon_{\gamma}-p_{t}\mathrm{g}_{\chi\gamma}+(p_{r}-p_{t})\xi_{\chi}\xi_{\gamma}
+\frac{1}{4 \pi} (-\mathcal{F}^{\zeta \chi} \mathcal{F}_{\eta \chi} + \frac{1}{4} \delta^{\zeta}_{\eta} \mathcal{F}^{\chi \psi} \mathcal{F}_{\chi \psi} ),
\end{equation}
where $p_{r}$ gives here the radial pressure, $p_{t}$ being the transversal pressure, and  the $\rho$ expresses the energy density. The term $\mathcal{F}^{\zeta \chi}$ exhibits the standard Maxwell's pressure tensor.
The 4-velocity vector is determined by $\upsilon_{\chi}$ and the expression $\xi_{\alpha}$ represents the radial 4-vector,with the following connections
\begin{equation}
\upsilon^{\alpha}=e^{\frac{-\nu}{2}}\delta^{\alpha}_{0},~~~\upsilon^{\alpha}\upsilon_{\alpha}=1,~~~\xi^{\alpha}
=e^{\frac{-\lambda}{2}}\delta^{\alpha}_{1},~~~\xi^{\alpha}\xi_{\alpha}=-1.
\end{equation}
The Einstein-Maxwell equations when assumed with the corresponding gravitational units, are read as
\begin{eqnarray}\nonumber
R_{\mu\nu}-\frac{1}{2} Rg_{\mu\nu} &=& -8\pi T_{\mu\nu},\\\nonumber
\mathcal{F}^{\eta\zeta}_{;\zeta}&=&-4 \pi j^\eta,\\\label{600}
\mathcal{F}_{[\beta\zeta;\eta]}&=&0,
\end{eqnarray}
where $j^\eta$ gives electromagnetic current 4-current vector. The following expressions respectively define the Maxwell Stress Tensor and the electromagnetic current 4-vector as
\begin{eqnarray}\nonumber
\mathcal{F}^{\eta \zeta}&=&A_{\zeta,\eta}-A_{\eta, \zeta},\\\label{12}
j^\eta&=&\sigma \nu^{\eta}.
\end{eqnarray}
Here, the letter $A$ corresponds to the magnetic 4-potential and the symbol $\sigma$ stands for the density.
The only non vanishing component in the static spherically symmetric system is $J^{0}$. The non-zero component $\mathcal{F}^{01}$  is merely attained by the Einstein-Maxwell vector and is given as
\begin{eqnarray}\label{13}
\mathcal{F}^{01}=-\mathcal{F}^{10}=\frac{q}{r^2} e^{-(\frac{\nu+\lambda}{2})},
\end{eqnarray}
with the letter $q$ being the stellar interior charge, defined as
\begin{eqnarray}\label{14}
q=4\pi\int_{0}^{r}\sigma \rho^2 e^{(\frac{\lambda}{2})} d\rho.
\end{eqnarray}
The term electric field intensity $E$ can be expressed as
\begin{eqnarray}
~~~~~~~~~~E^2=-\mathcal{F}^{01}\mathcal{F}_{10}=\frac{q^2}{ r^4}.~~~~~~~~~~\quad\quad~~~~~~\label{19}
\end{eqnarray}
Now employing the Schwarzschild exterior space-time, i.e., $e^{\nu(r)}=e^{-\lambda(r)}=\frac{r-2M}{r}$, in Eq. (\ref{1}), being embedded into $5-D$
Euclidean space-time, we employ the transformation
\begin{equation}\label{2}
\mathbbm{x}=r sin \theta cos \phi, \;\;\;\;\;\;\mathbbm{y}=r sin \theta sin \phi,\;\;\;\;\;\;\;\mathbbm{z}=r cos\;\theta
\end{equation}
Now, Eq. (\ref{1}) gets converted into the following space-time:
\begin{equation}\label{3}
 d{s}^2=\frac{r-2M}{r}dt^2-\frac{2M}{r-2M}d{r}^2-d\mathbbm{x}^{2}-d\mathbbm{y}^{2}-d\mathbbm{z}^{2},
\end{equation}
where $r=\sqrt{\mathbbm{x}^{2}+\mathbbm{y}^{2}+\mathbbm{z}^{2}}$. By considering the following relations:
\begin{equation}\label{4}
 dR_b=\sqrt{\frac{2M}{r-2M}}dr,\;\;\;\;\;\;\;\;\;\;R_b=\sqrt{8M(r-2M)},
\end{equation}
Eq. (\ref{3}), becomes
\begin{equation}\label{5}
 d{s}^2=-d{R_b}^2-d\mathbbm{x}^{2}-d\mathbbm{y}^{2}-d\mathbbm{z}^{2}+\frac{R_b^{2}}{16M^{2}+R_b^{2}}dt^2.
\end{equation}
Here, the coefficient of last term in above equation is not provided the proper differential form, this concept leads that the line element from Eq. (\ref{1}) is not embedded to 5-D space-time. Therefore, we need to define another transformation:
\begin{equation}\label{6}
 d{S}^2=\frac{R_b^{2}}{16M^{2}+R_b^{2}}dt^2-d{R_b}^2=d\mathbbm{x_{1}}^{2}+d\mathbbm{y_{1}}^{2}+d\mathbbm{z_{1}}^{2},
\end{equation}
where,
\begin{equation}\label{7}
\mathbbm{x_{1}}=\frac{R_b \times sin t}{(16M^{2}+R_b^{2})^{\frac{1}{2}}}, \;\;\;\mathbbm{y_{1}}=\frac{R_b\times cos t}{(16M^{2}+R_b^{2})^{\frac{1}{2}}},\;\;\;\mathbbm{z_{1}}=\int\left(\frac{(16M^{2}+R_b^{2})^{3}+
256M^{4}}{(16M^{2}+R_b^{2})^{3}}\right)^{\frac{1}{2}}dR_b.
\end{equation}
Now, finally Eq. (\ref{5}), provides a embedded 6-D, Euclidean space-time, which is calculated as:
\begin{equation}\label{8}
 d{s}^2=-d\mathbbm{x}^{2}-d\mathbbm{y}^{2}-d\mathbbm{z}^{2}+ d\mathbbm{x_{1}}^{2}+d\mathbbm{y_{1}}^{2}+d\mathbbm{z_{1}}^{2}.
\end{equation}
In 1975, Gupta et al. \cite{5901} also provided a 6-D embedded Euclidean space-time, which is calculated below:
\begin{equation}\label{9}
 d{s}^2= d\mathbbm{x_{1}}^{2}+d\mathbbm{y_{1}}^{2}\pm d\mathbbm{z_{1}}^{2}-d\mathbbm{x}^{2}-d\mathbbm{y}^{2}-d\mathbbm{z}^{2},
\end{equation}
where,
\begin{eqnarray*}
\mathbbm{x}&&=r sin \theta cos \phi,\;\;\;\;\; \;\;\;\;\;\;\;\;\;\mathbbm{y}=r sin \theta sin \phi,\;\;\;\;\;\;\;\;\;\;\;\;\;\;\;\mathbbm{z}=r cos\;\theta,\nonumber\\
\mathbbm{X}&&=K_{1}e^{\frac{\nu(r)}{2}} cosh(\frac{t}{K_{1}}), \;\;\;\mathbbm{Y}=K_{1}e^{\frac{\nu(r)}{2}} sinh(\frac{t}{K_{1}}),\;\;\;\mathbbm{Z}=H(r).
\end{eqnarray*}
Here, $K_{1}>0$, is unknown parameter, and $H^{'^{2}}(r)=\mp(-(e^{\lambda(r)}-1)+\frac{K_{1}^{2}e^{\nu(r)}\nu^{'^{2}}(r)}{4}$.\\
Now we can calculate the gravitational potentials from the last transformation, under the assumption $H^{'^{2}}(r)=0$, as
\begin{equation}\label{17}
e^{\lambda(r)}=1+L\times e^{\nu(r)}\nu^{'^{2}}(r).
\end{equation}
where $L=\frac{K_{1}^{2}}{4}\neq 0$ is an arbitrary constant. This is well known ``Karmarker condition'' \cite{76}, involving the metric coefficients.
In fact, it is an analytical approach to deal with the Einstein field equations and investigate the physically stable models. This condition has been proved to be an important mechanism to explore new solutions of astrophysical models.
Moreover, the theoretical possibility of studying stellar models in the case of anisotropic and charged distributions has been done previously by many authors, for some references see (\cite{BBB}-\cite{Takisa}). For massive stellar objects, the radial pressure may not be equal to the tangential one. Some important arguments are witnessed for the existence of anisotropy in stellar models such as by the presence of type $3$A superfluid \cite{8143} and different kinds of phase transitions \cite{9143}. As far as the inclusion of charge is concerned,
Rosseland \cite{2600} studied the possibility of a self gravitating star
treated as a ball of hot ionized gases having a considerable amount of charge.
A large supply of electrons in such a system as compared to positive ions, run to
escape from its surface due to their higher kinetic energy and the motion of electrons remain continued until the electric field is present inside the compact star. In this way, the equilibrium is attained
after some electrons escape and the net electric charge approaches to
about 100 Coulombs per solar mass. Thus, the possibility of collapsing of stellar structure to a point
singularity may be avoided by the effects of charge. So, in this work we are motivated to investigate the charged compact star solutions
by using the Karmarkar condition, in particular, with Bardeen black hole geometry.
For the sake of simplicity, from now onwards we consider $\lambda(r)\equiv \lambda,~\nu(r)\equiv \nu$.

Now for the geometry (\ref{1}) in the presence of anisotropic content (\ref{4002}), equations (\ref{600}) have the new expressions as
\begin{eqnarray}\label{F1}
8\pi\rho+E^2&=&\frac{1}{ e^{\lambda}r^2}\big(e^{\lambda}+\lambda'r-1\big),
\\\label{F2}
8\pi p_r-E^2&=&\frac{1}{ e^{\lambda}r^2}\big(-e^{\lambda}+\nu'r+1\bigg),~\\\label{F3}
8\pi p_t+E^2&=&\frac{1}{ e^{\lambda}}\big(\frac{{\nu'}^{2}}{4}+\frac{\nu''}{2}-\frac{\nu'\lambda'}{4}+\frac{\nu'}{2r}-\frac{\lambda'}{2r}\big),\\\label{F4}
 \sigma&=& \frac{e^{-\lambda/2}}{4\pi r^2}(r^2E)'.
\end{eqnarray}
Now, our next task is to find the solution of the Einstein-Maxwell Eqs. (\ref{F1})-(\ref{F4}) with the manipulation of Eq. (\ref{17}). We acquire here five equations with the six unknowns namely $\nu,~ \lambda,~ \rho,~ p_r,~ p_t$ and $E$. Here in order to find a solution, we ansatz the metric time co-efficient as proposed by Adler \cite{Adler}
\begin{equation}\label{adler}
e^\nu = X(1 + Yr^2)^2,
\end{equation}
where $X$ and $Y$ being the arbitrary non-zero constants. This is important choice as in the low mass limit the solution becomes identical to the Schwarzschild interior solution.
It may be worth noticing here that some fascinating results have already been reported using exactly similar metric potential by Bhar et al. \cite{Bhar}. Now by inducting an electric field of the form $E^2 =KYr$.
and manipulating the equation (\ref{17}) and(\ref{adler}), it follows
\begin{equation}\label{finch}
e^\lambda = 1 + 16XY^2L r^2,
\end{equation}
which is similar to one given in Finch-Skea
solution \cite{Finch}. Now engaging the metric potentials from
equations (\ref{adler}) and (\ref{finch}) with the Einstein-Maxwell’s field equations (\ref{F1})-(\ref{F4}) and employing the expression for $E^2$, we get the expression for the physical quantities as
\begin{eqnarray}\nonumber
\rho&=&\frac{1}{8\pi\left(16 X Y^2 L r^2+1\right)^2}\times(-256 X^2 Y^5 L^2 K r^5+256 X^2 Y^4 L^2 r^2\nonumber\\&&-32 X Y^3 L K r^3+48 X Y^2 L-Y K r),
\label{F11}\\
p_r&=&\frac{1}{8\pi\left(Y r^2+1\right) \left(16 X Y^2 L r^2+1\right)}\times(16 X Y^4 L K r^5+16 X Y^3 L K r^3\nonumber\\&&-16 X Y^3 L r^2-16 X Y^2 L+Y^2 K r^3+Y K r+4 Y),\label{F22}\\
p_t&=&\frac{1}{8\pi\left(Y r^2+1\right) \left(16 X Y^2 L r^2+1\right)^2}\times(-256 X^2 Y^6 L^2 K r^7\nonumber\\&&-256 X^2 Y^5 L^2 K r^5-32 X Y^4 L K r^5-32 X Y^3 L K r^3+16 X Y^3 L r^2\nonumber\\&&-16 X Y^2 L-Y^2 K r^3-Y K r+4 Y),\label{F33}\\
 \sigma&=& \frac{5 Y K}{8 \pi  \sqrt{Y K r(16 X Y^2 L r^2+1)}}.\label{F44}
\end{eqnarray}
The anisotropy parameter $\triangle$ turns out to be
\begin{eqnarray}\nonumber
\triangle=p_t-p_r&=&\frac{1}{4 \pi  \left(Y r^2+1\right) \left(16 X Y^2 L r^2+1\right)^2}\times (-256 X^2 Y^6 L^2 K r^7-
256\nonumber\\&&\times X^2 Y^5 L^2 K r^5+128 X^2 Y^5 L^2 r^4+128 X^2 Y^4 L^2 r^2-32 X Y^4 L\nonumber\\&&\times K r^5-32 X Y^3 L K r^3-16 X Y^3 L r^2-Y^2 K r^3-Y K r).\label{anisotropy}
\end{eqnarray}
Now an important discussion concerning the suitable boundary constraints for the matching purpose with the interior spacetime is provided in the next section.

\section{Matching Conditions}

The following Bardeen spacetime is proposed for the  description of the exterior stellar
geometry \cite{Bardeen}, that is
\begin{equation}\label{bardeen}
ds^{2}={f(r)}^{-1}dr^2+r^{2}d\theta^{2}+r^2sin^{2}\theta d\phi^{2}-f(r)dt^{2},
\end{equation}
where $f(r)=1 -\frac{2Mr^2}{({q^2}+{r^2})^{\frac{3}{2}}}$.
It may be worth noticing here that the Bardeen black hole geometry may be described as the magnetic monopole of gravitationally collapsing remnants emerging from particular case of some non-linear electrodynamics \cite{Garcia}. Also, the Bardeen black holes structures may be taken as the analytic solutions of some proper non-linear electrodynamics connected to the gravity. In additon to this, the non-zero components of Einstein tensor in the Bardeen geometry can be related to the stress-energy tensor of the Lagrangian of nonlinear electromagnetic field \cite{Moreno}. In addition to this, the Bardeen structures are not in contradiction to existing singularity theorems \cite{Hawking} in the literature. Studying Bardeen black hole geometries emerged as a matter of vital importance to the researcher in the past \cite{Fernando1}-\cite{Ulhoa}. It should be noted here that the spacetime exhibits the asymptotically behavior as
\begin{equation}\label{22}
f(r) = -\frac{2M}{r}+\frac{3Mq^2}{r^3}+1+ O(\frac{1}{r^5}).
\end{equation}
The term $r^{-1}$ appearing in Eq. (\ref{22}) hints that the parameter $M$ is correlated to the stellar mass. Though, the next term containing $r^{-3}$ turns the situation even more mesmerizing as it does not involve the charge parameter $q$ (Coulomb) similar to the case of Reissner–Nordstrom solution \cite{Nordstrom}. Therefore, for our ongoing investigations, we consider $ 1-\frac{2M}{r}+\frac{3Mq^2}{r^3}\approx f(r) $. Having inspired from the developing situation as discussed, incorporating  Bardeen geometry to investigate the compact structures could certainly prove to be quite effective. Particularly the engrossment of $\frac{3Mq^2}{r^3}$ in the Bardeen geometry relative to the term $\frac{q^2}{r^2}$ into the existing Reissner Nordstrom geometry may deliver some enthralling outcomes. Darmois \cite{Darmois4} and Israel \cite{Israel2} proposed a smooth evaluation between the inner and the outer regions. Using the continuity of the metric potentials on the boundary, we obtain the following matching equations
\begin{eqnarray}\label{M1}
1-\frac{2M}{{R_b}}+\frac{3Mq^2}{{R_b}^3}&=&X(1 + Y{R_b}^2)^2,\\\label{M2}
(1-\frac{2M}{{R_b}}+\frac{3Mq^2}{{R_b}^3})^{-1}&=&1 + 16XY^2L {R_b}^2,\\\label{M3}
p_r(r = R_b) &=& 0.
\end{eqnarray}
Now, with the implication of the imposed comparing expressions (\ref{M1})-(\ref{M3}), we read
\begin{eqnarray}\nonumber
L&=&\frac{1}{64 M (M+{R_b})^2}\times \big[420 M^3 {R_b}^2 + 44 M^2 {R_b}^3 + 216 K M^3 {R_b}^3 + 32 M {R_b}^4
\nonumber \\&&+144 K M^2 {R_b}^4 +
 27 K^2 M^3 {R_b}^4 + 16 {R_b}^5 +48 K M {R_b}^5 + 27 K^2 M^2 {R_b}^5\nonumber \\ &&+ 8 K {R_b}^6 +
 9 K^2 M {R_b}^6 + K^2 {R_b}^7+(42 M^2 + 18 M {R_b} + 9 K M^2 {R_b} +\nonumber\\&& 4 {R_b}^2 + 6 K M {R_b}^2 + K {R_b}^3)\times\bigg({R_b}^4 \left(M^2 (3 K {R_b}+10)^2+2 M {R_b} (3 K \right.\nonumber\\&&\times\left.{R_b} (K {R_b}+2)-40)+{R_b}^2 (K {R_b}+4)^2\bigg)^{\frac{1}{2}}\right)~\big],\label{M11}
\end{eqnarray}
\begin{eqnarray}\nonumber
X&=&\frac{1}{32 {R_b}^3 (3 K M {R_b}+2 M-{R_b})}\times \big[{R_b}^2(-45 K^2 M^2 {R_b}^2-18 K^2 M {R_b}^3\nonumber \\&&-K^2 {R_b}^4-180 K M^2 {R_b}+44 K M {R_b}^2-100 M^2+80 M {R_b}-16 {R_b}^2)\nonumber\\
&&+(15 K M {R_b}+K {R_b}^2+10 M-4 {R_b})(2 KY {R_b}^5 (15 M+{R_b})+{R_b}^3\nonumber\\
&&\times (3 K M+4)+K {R_b}^4-10 M {R_b}^2)\big],\label{M22}\\
Y&=&\frac{1}{2 K {R_b}^5 (15 M+{R_b})}\times \big[10 M {R_b}^2-{R_b}^3 (3 K M+4)-K {R_b}^4+\nonumber\\&&\bigg({R_b}^4 \left((M (3 K {R_b}-10)+{R_b} (K {R_b}+4))^2+8 K\right.\nonumber\\
&&\times\left. M {R_b} (15 M+{R_b})\right)\bigg)^{\frac{1}{2}}~\big].\label{M33}
\end{eqnarray}
In fact, $L,~ X$ and $Y$ are the model parameters which are obtained using the continuity conditions over the boundary of star. All these depend upon some free parameters $R_b,~ M$ and $K$. We have chosen the value of $K=.002$ to obtain corresponding values (kindly see Table-\textbf{I}), which will provide some good physical behavior eventually. Now, we provide a critical analysis via the independent parameters $R_b,~ M$ and $K$ appearing in the expressions (\ref{M11})-(\ref{M33}).
\begin{center}
\begin{table}
\caption{\label{tab1}{Approximated values of $Y,\; X,$ and $L$ with $K=0.002$.}}
\begin{tabular}{|c|c|c|c|c|c|c|c|c|}
    \hline
    \multicolumn{4}{|c|}{$R_{b}=10.10 (km)$} \\
    \hline
$M(M_{\odot})$ \;\;\;\;                  & $Y (km^{-2})$\;\;\;\                  &$X (km^{-2})$\;\;\;               &$L(km^{-1})$\\
\hline
2.00\;\;\;\;                             &0.0018779264054689           &0.4269934485075686      &264.2475639283101 \\

2.20\;\;\;\;                             &0.0022773989755877           &0.3736472113871004      &241.0203567212078\\

2.40\;\;\;\;                             &0.0027659091165267           &0.3216815781294093      &221.8309760098252\\
    \hline
    \multicolumn{4}{|c|}{$R_{b}=09.10 (km)$} \\
    \hline
2.00\;\;\;\;                             &0.0028677203968403           &0.3678391597906813      &193.4318750201828 \\

2.20\;\;\;\;                             &0.0035631904817850           &0.3102660507012401      &176.5892849189401\\

2.40\;\;\;\;                             &0.0044588601352241           &0.2548757844016362      &162.7549546837323\\
    \hline
    \multicolumn{4}{|c|}{$R_{b}=08.10 (km)$} \\
    \hline
2.00\;\;\;\;                             &0.0047551973225106           &0.2962369192451133      &136.6798131264395\\

2.20\;\;\;\;                             &0.0061487590251187           &0.2346253568229378      &124.9968261714293\\

2.40\;\;\;\;                             &0.0081038873167639           &0.1769170571727629      &115.5439133984426\\
\hline
\end{tabular}
\end{table}
\end{center}

\section{Physical Analysis}

The detail analysis of current study with analytic and graphic discussion is provided in this section.

\subsection{Evolution of metric functions}

The gravitational components, i.e., $g_{rr}$, and $g_{tt}$, of the spherically symmetric space-time both are seen physically acceptable with the condition, i.e., $g_{rr}(r =0)=1$ and $g_{tt}(r =0)\neq0$. The variational character of both components of the metric may be revealed as reflected in right panel of the left part of Fig. \textbf{1}, against the three different values of $M$ for three various ranges dependent on the radial coordinate $r$, i.e., $r=R_{b}=10.10,\;r=R_{b}=09.10$ and $r=R_{b}=08.10$. In all cases both the metric functions satisfy the basic criterion for configuration of compact star.

\subsection{Energy density}

The right part of Fig. \textbf{1} discusses the graphic development of energy density function of this current study. From same plot, it is cleared that energy density function is observed positive against the three different values of $M$. It is also observed that energy density at center, i.e., $\rho(r=0)$ is seen maximum and minimum at boundary, i.e., $\rho(r=R_{b})$. This decreasing attribute of energy density function shows the stability of configuration of compact stars.
\begin{figure}
\centering\epsfig{file=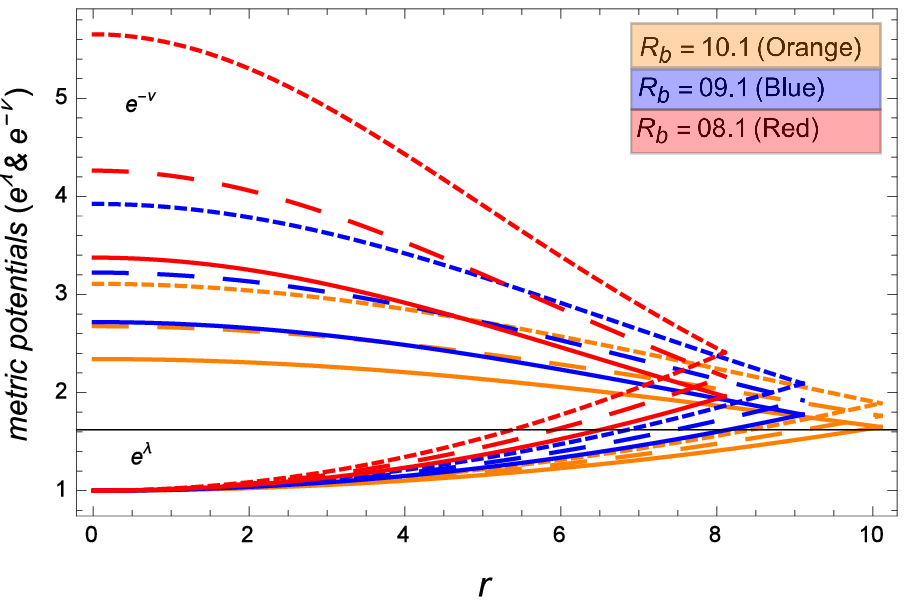, width=.48\linewidth,
height=2.3in}\epsfig{file=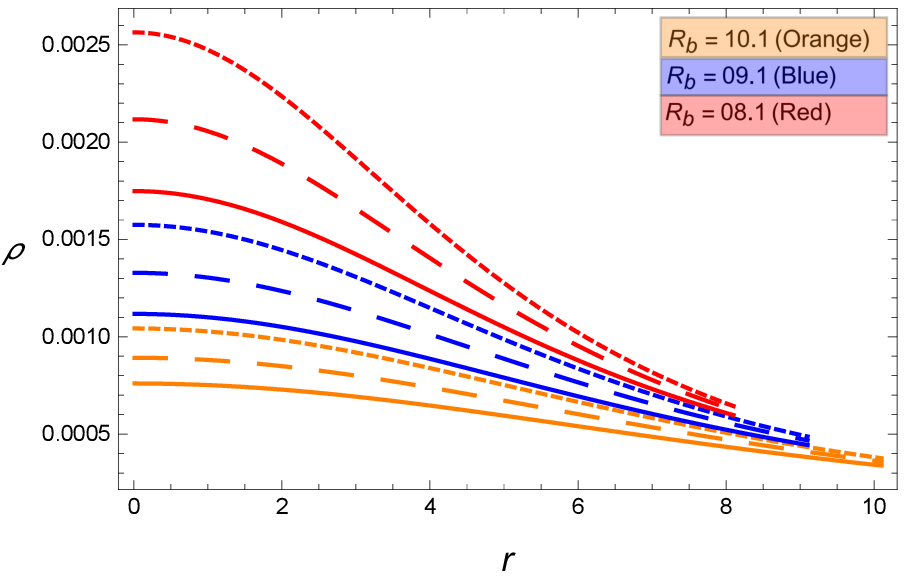, width=.48\linewidth,
height=2.3in}\caption{\label{Fig.2} Reflects the evolving profile of metric functions and energy density}
\end{figure}

\subsection{Pressure Components}

The pressure distribution in the anisotropic background can be divided mainly into two portions, acknowledged as tangential and radial pressure. As the role of pressure and energy density is eminent in the evolution of the  compact objects, therefore, their analysis in our study is important as well. From Fig. \textbf{2}, it can be seen that both of the pressure sources exhibit the increasing trend. The tangential pressure thoroughly remains positive for the diverse choices of the radial ranges, i.e., $0<r\leq 8.10,\;0<r\leq 9.10 $ \& $ 0<r\leq 10.10$ by considering $M$ as $M=2.00,~\;M=2.20 $ \& $ M=2.40$. It is also evident that at the boundary, the radial pressure vanishes, i.e., $r=R_{b}$. Both of the pressure terms are reported maximum at the center, i., $r=0$, and both with charge field meet the desired condition of the compact objects.

\subsection{Charge density and electric field}

It may be noticed from the evolution of Fig. \textbf{3} that the electric field turns zero at center and then increases with positive nature outward, and the charge density decreases from center to boundary. This is an ideal behavior which in fact supports the other physical parameters to be well-behaved. For example, the energy density function and pressure components show much more realistic trends due to the involvement of charge in the fluid.
\begin{figure}
\centering \epsfig{file=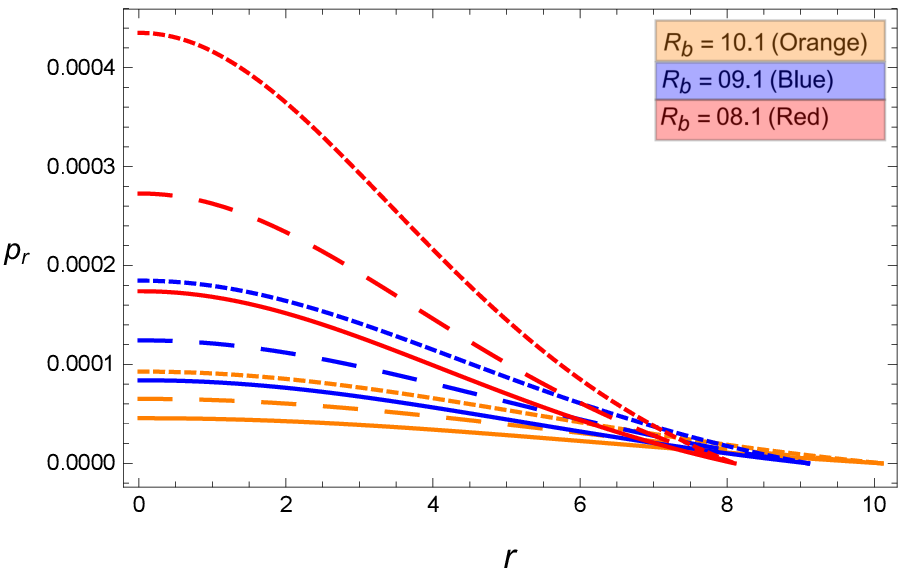, width=.48\linewidth,
height=2.3in}\epsfig{file=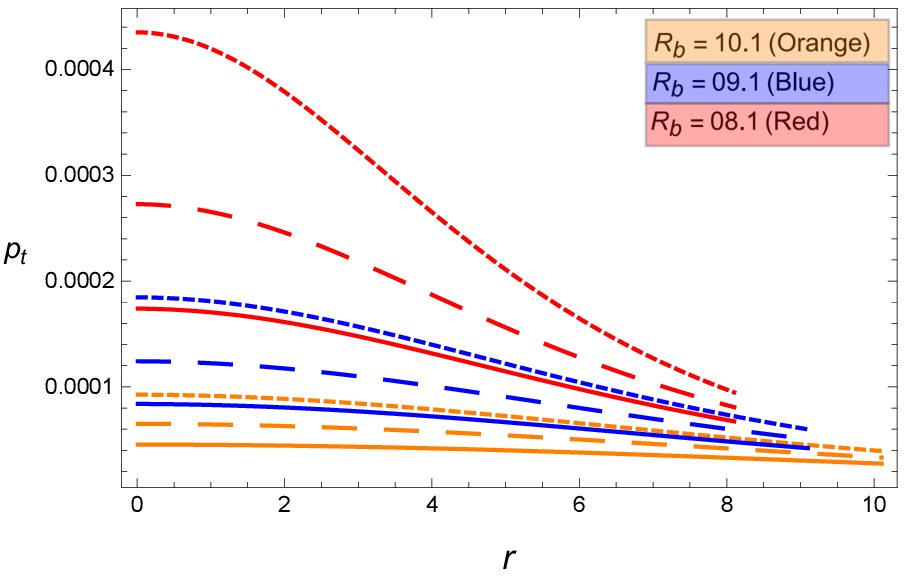, width=.48\linewidth,
height=2.3in}\caption{\label{Fig.2} Reflects the evolving profile of the pressure components}
\end{figure}
\begin{figure}
\centering \epsfig{file=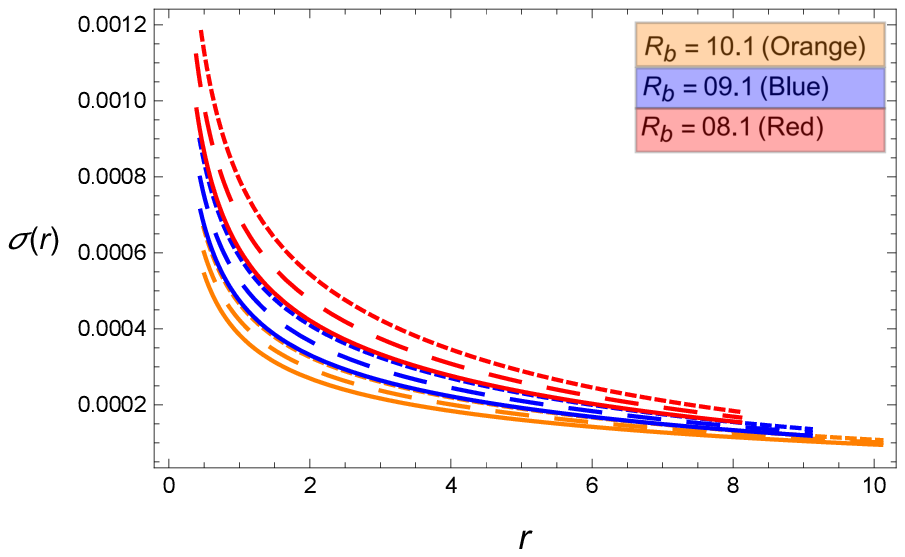, width=.48\linewidth,
height=2.3in}\epsfig{file=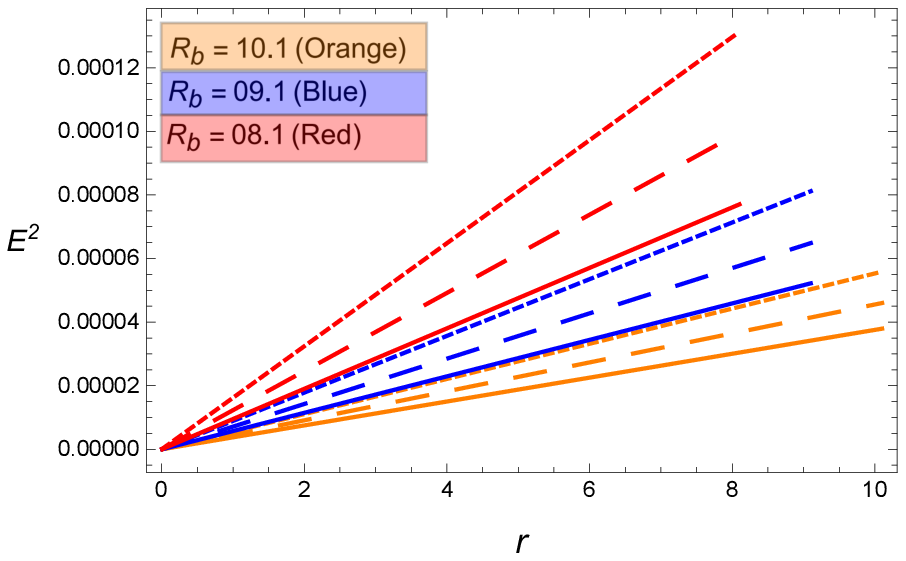, width=.48\linewidth,
height=2.3in}\caption{\label{Fig.3} Reflects the evolving profile of the behavior of charge density and the electric field}
\end{figure}

\subsection{Anisotropy}

The expression defined and represented by $\triangle=p_{t}-p_{r}$ provides a concept of anisotropic function. It may be perceived from the plots of Fig. \textbf{2} that both the radial and tangential pressure are seen equal at $r=0$, and then both the pressure components vary from center to boundary, the tangential component remains positive while the radial pressure vanishes at the boundary, this idea leads that anisotropic function always remains positive throughout the configuration. The graphical nature of anisotropic function can be checked from Fig. \textbf{4}, which is observed positive with increasing trend. \begin{figure}
\centering \epsfig{file=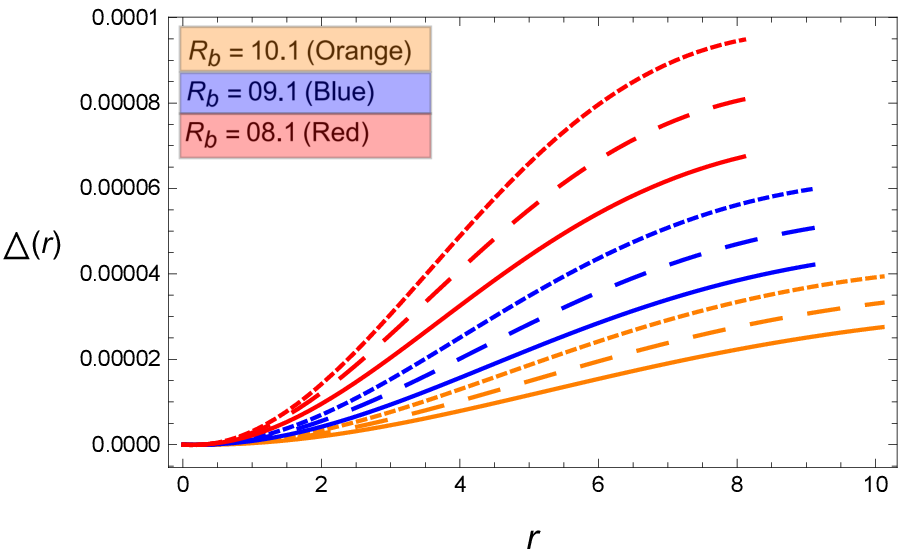, width=.48\linewidth,
height=2.3in}\epsfig{file=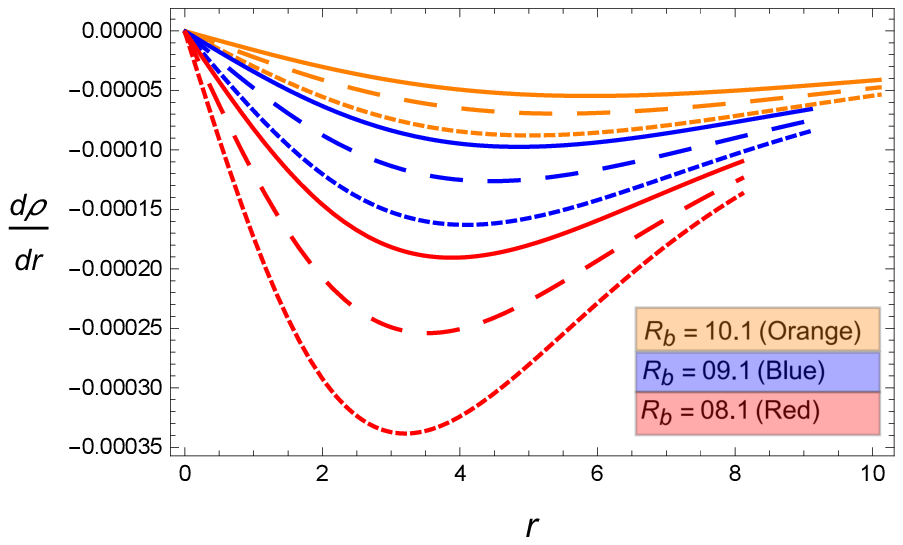, width=.48\linewidth,
height=2.3in}
\caption{\label{Fig.4} Reflects the evolving profile of the behavior of anisotropy function and energy gradient}
\end{figure}

\subsection{Non-singular nature of current study}

Herein, we shall discuss about Zeldovich's condition and other physical parameters at $r=0$.
\begin{eqnarray*}
\rho_{0}&=&\rho(r =0)=\frac{6 X Y^2 L}{\pi },\label{32}\\
p_{r0}&=&p_{r}(r =0)=\frac{4 Y-16 X Y^2 L}{8 \pi },\label{33}\\
p_{r0}&=&p_{r}(r =0)=p_{t0}=p_{t}(r =0).\label{34}
\end{eqnarray*}
From above the calculations, the ratio $p_{r0}/ \rho_{0}=p_{t0}/ \rho_{0}$, is satisfied obeying the Zeldovich's condition and calculated as
\begin{equation}
\frac{p_{r0}}{\rho_{0}}=\frac{1}{12} \left(\frac{1}{X Y L}-4\right)\leq1
\end{equation}

\subsection{Gradients}

The gradients of energy density, and both of the pressure terms with respect to $r$, for the current study are shown in Figs. \textbf{4-5}. It is evident that
\begin{equation*}
0>\frac{d\rho}{dr},\;\;\;\;\;\;\; \;0>\frac{dp_{r}}{dr},\;\;\;\;\;\;\;0>\frac{dp_{t}}{dr}.
\end{equation*}
The negative evolution of the gradients is considered a necessary condition for the compact stars configuration. However,
\begin{equation*}
\frac{d\rho}{dr}(r=0)= 0,\;\;\;\;\;\;\; \;\frac{dp_{r}}{dr}(r=0)= 0,\;\;\;\;\;\;\;\frac{dp_{t}}{dr}(r=0)=0
\end{equation*}
The above relations show that gradients vanish at $r=0$.
\begin{figure}
\centering \epsfig{file=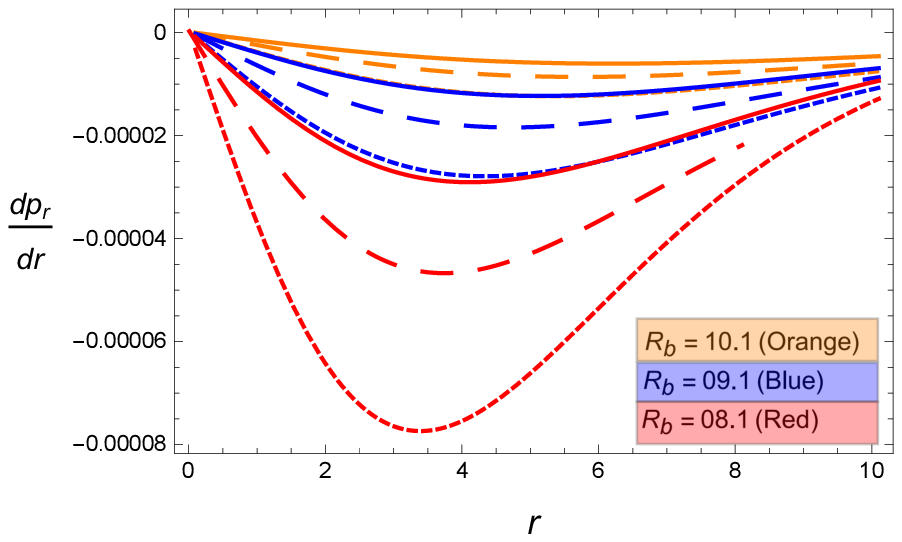, width=.48\linewidth,
height=2.3in}\epsfig{file=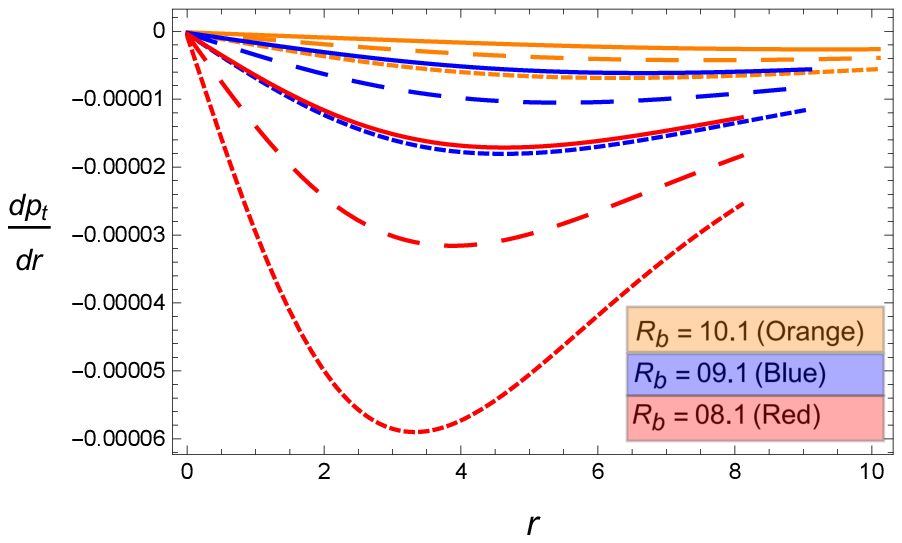, width=.48\linewidth,
height=2.3in}\caption{\label{Fig.5} Reflects the evolving profile of pressure gradients}
\end{figure}

\subsection{Energy Conditions}
Energy constraints play a vital role in general relativity. The following four constraints are described as null energy condition $NEC$, strong energy condition $SEC$, weak energy condition $WEC$, and dominant energy condition $DEC$, which are defined as
\begin{eqnarray}\nonumber
NEC: \quad&& \rho+p_{t}\geq0, \quad \rho+p_{r}\geq0,\;\;\;\;\;\;\;\;\; \;\;\;\;\;\;\;\;\;\;\;\;\;\;\;WEC:\quad 0\leq\rho,\quad\rho+p_{t}\geq0, \quad \rho+p_{r}\geq0,\nonumber\\
SEC: \quad&& \rho-p_{r}-2p_{t}\geq0,\;\;\;\;\;\;\;~~~~~~~~~~~~~~~~~~~~~~~~DEC: \quad 0\leq\rho,\quad\rho>|p_r|, \quad\rho>|p_t|.\nonumber
\end{eqnarray}
\begin{figure}
\centering \epsfig{file=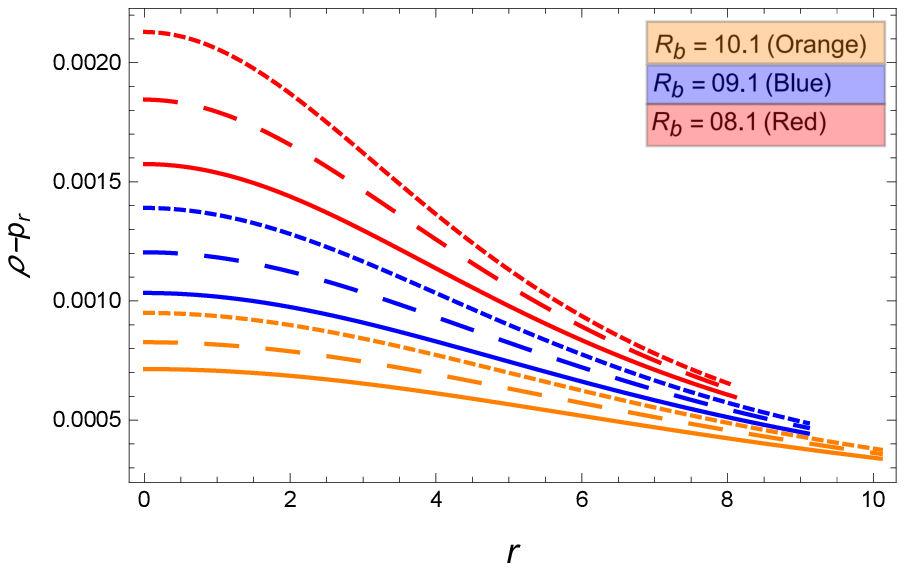, width=.48\linewidth,
height=2.3in}\epsfig{file=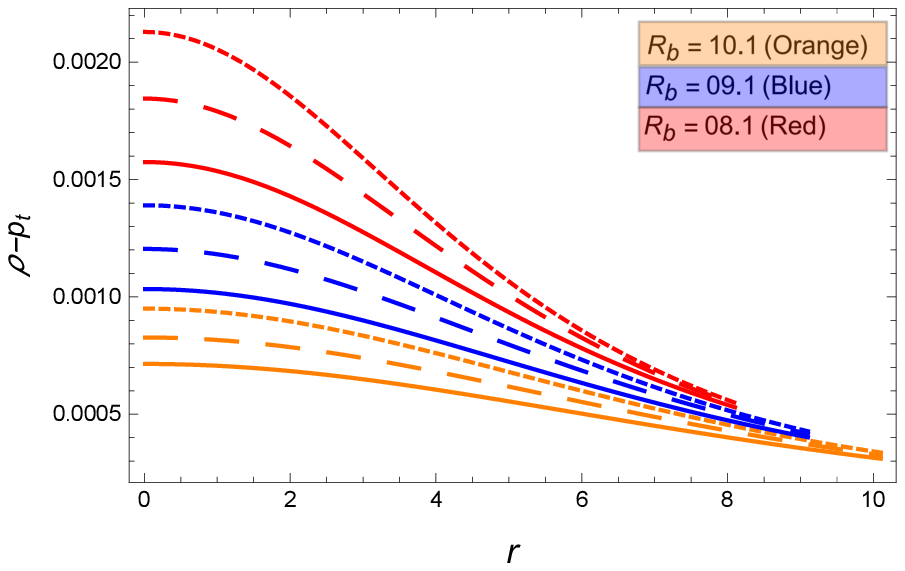, width=.48\linewidth,
height=2.3in}\caption{\label{Fig.6} Reflects the evolving profile of DEC}
\end{figure}
\begin{figure}
\centering  \epsfig{file=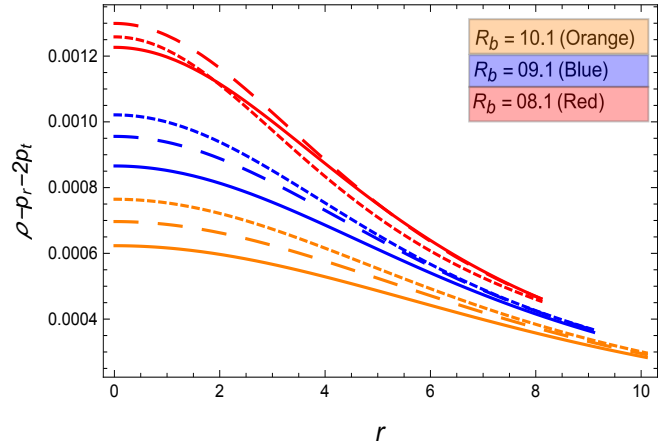, width=.48\linewidth,
height=2.3in}\caption{\label{Fig.7} Reflects the evolving profile of SEC}
\end{figure}
The $DEC$ and $NEC$ are justified for our study and can be seen from Figs. \textbf{1-2}. The $SEC$ and $WEC$ are also fulfilled for the present study as depicted from Figs. \textbf{6-7}. All the energy bounds are seen satisfied, which is another positive sign for the existence of charged compact stars for the Bardeen model in a realistic way.
\subsection{ Equation of state}
Two important ratios $\frac{p_r}{\rho}$ and $\frac{p_t}{\rho}$ provide the basic idea of equation of state parameters, i.e., $w_r$ and $w_t$, expressed as
\begin{equation}\label{31}
w_r\times \rho =p_r,\;\;\;\;\;\;\;\;\;\;\;\;\;\;\;\;\;\;\;\;w_t\times \rho =p_t.
\end{equation}
 In the analysis of current study, both the ratios, i.e., $\frac{p_r}{\rho}$ and $\frac{p_t}{\rho}$ remain less than 1. The $0<w_r $ \& $ w_t<1$ nature can be checked from the Fig. \textbf{8}. The $\frac{p_r}{\rho}$ ratio is seen maximum at $r=0$ and is the least at $r=R_{b}$. The $\frac{p_r}{\rho}$ ratio is noted minimum at $r=0$ and at $r=R_{b}$ it acquires the maximum value .
\begin{figure}
\centering \epsfig{file=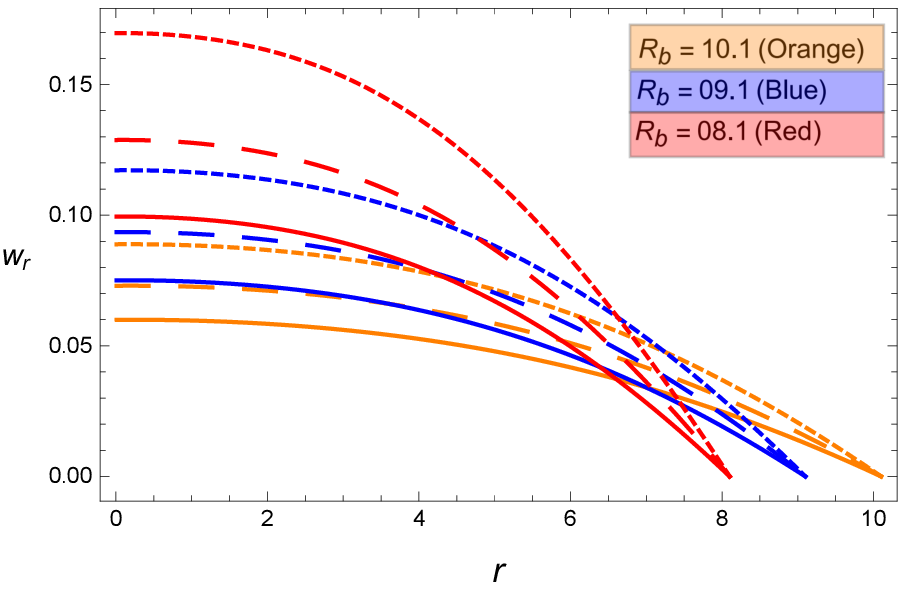, width=.48\linewidth,
height=2.3in}\epsfig{file=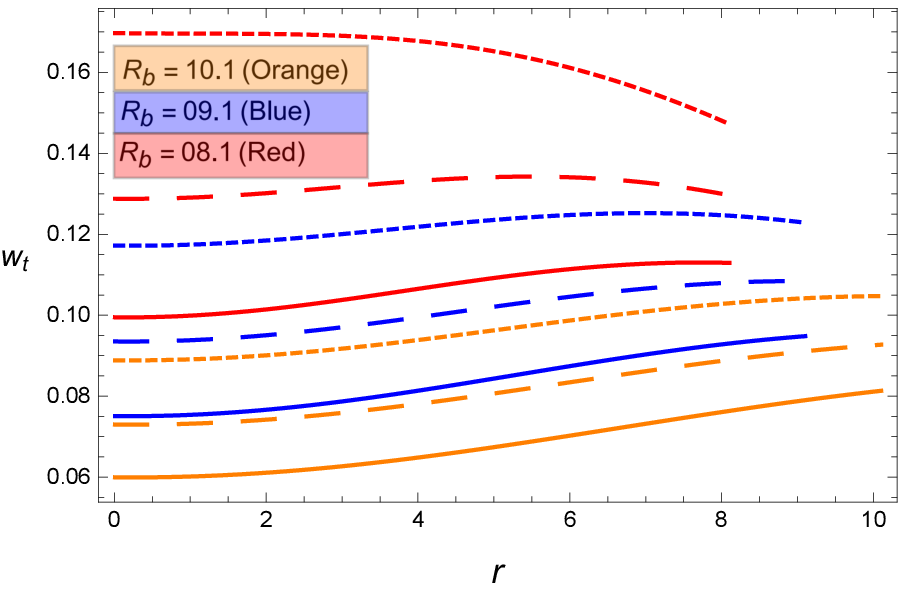, width=.48\linewidth,
height=2.3in} \caption{\label{Fig.8} Reflects the evolving profile of behavior of equation of state parameters}
\end{figure}

\subsection{Redshift, Compactness parameter, and the Mass function}

The redshift expression of the stellar structure is given as:
\begin{equation}\label{32}
Z(r)=\frac{1}{\sqrt{1-2u(r)}}\left(1-\sqrt{1-2u(r)}\right),
\end{equation}
where $u(r)$ mentions the compactness parameter, defined as:
\begin{equation}\label{33}
u(r)=\frac{2}{r}\times m(r),
\end{equation}
here $m(r)$ denotes the mass-function of compact stars and is determined as:
\begin{equation}\label{34}
m(r)=4\pi \times\int^{r}_{0}(r^{2}\times \rho)dr.
\end{equation}
For our work, the $m(r)$ is worked out as:
\begin{equation}\label{35}
m(r)=\frac{1}{8} r \left(-\frac{4}{16 X c^2 L r^2+1}-c K r^3+4\right)
\end{equation}
Manipulating Eq.(\ref{35}) and Eq. (\ref{33}), we acquire the compactness parameter, by using this current value of compactness parameter in Eq. (\ref{32}), we get the red-shift function for this current study.
\begin{figure}
\centering \epsfig{file=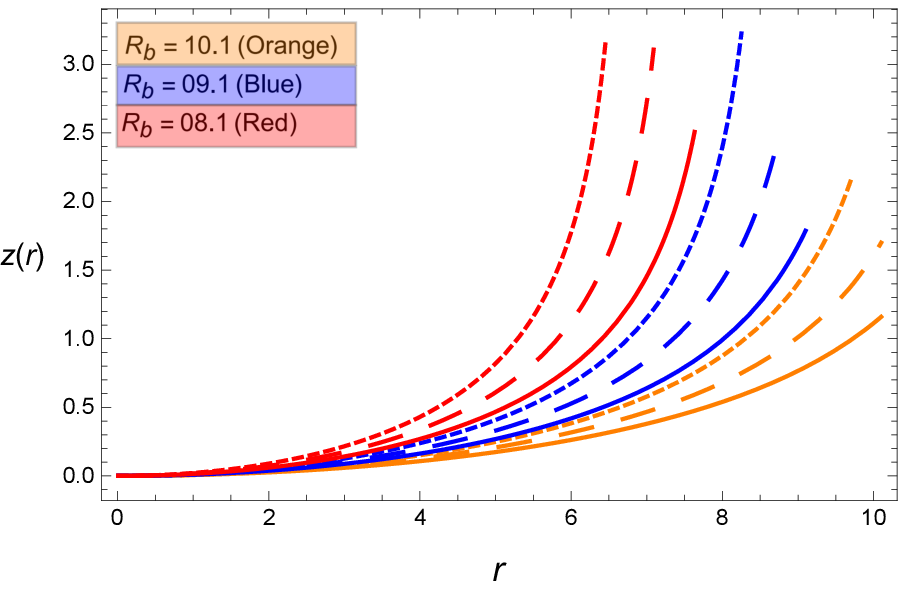, width=.48\linewidth,
height=2.3in}\epsfig{file=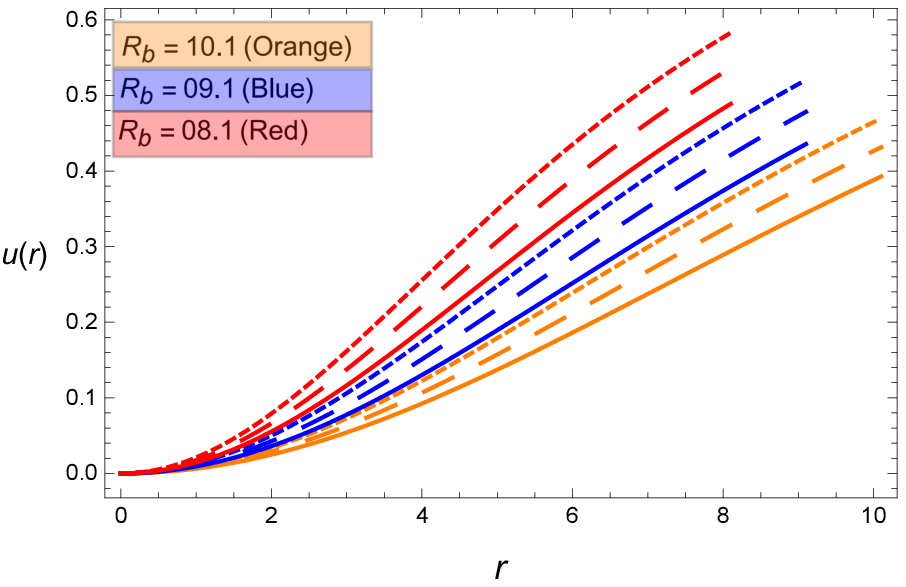, width=.48\linewidth,
height=2.3in} \caption{\label{Fig.9} Reflects the evolving profile of the behavior of red-shift function and compactness parameter}
\end{figure}
These above three physical parameters are very necessary for the compact stars study. The evolution of the red-shift parameter $Z(r)$, can be checked from Fig. \textbf{9} that satisfies the restriction of $Z(r)\leq 5$ \cite{40},and fulfilling the Ivanov condition \cite{41} of $Z_{s}\leq 5.211$ as well. In our analysis, $Z(r)\leq 3.100$, validates the physical acceptability of our study. The compactness $u(r)$ as evident from Fig. \textbf{9}, also justifies the Buchdahl condition \cite{42}. The function $m(r)$, is plotted in Fig. \textbf{10}. It is easy to see from the same Fig. that the $m(r)$ is increasing monotonically towards the surface.

\subsection{Equilibrium Condition}

Now, we discuss hydrostatic-equilibrium configuration for our study. Therefore,we discuss the Tolman-Oppenheimer-Volkoff (TOV) equation with electric field. This equation is calculated after solving the Einstein conservation equation in the presence of electric charge. For the anisotropic content, modified TOV is defined as:
\begin{equation}\label{36}
\frac{2}{r}(p_{t}-p_{r})-\frac{d p _{r}}{dr}-\frac{\nu(r)'}{2}(\rho+p_{r})+ \sigma(r)E(r)e^{\frac{\lambda (r)}{2}}=0,
\end{equation}
The above equation can be converted as:
\begin{equation}\label{37}
\mathscr{F}_{\mathrm{h}}+\mathscr{F}_{\mathrm{g}}+\mathscr{F}_{\mathrm{a}}+\mathscr{F}_{\mathrm{e}}=0,
\end{equation}
where
\begin{equation*}
\mathscr{F}_{\mathrm{a}}=\frac{2}{r}(p_{t}-p_{r}),\;\mathscr{F}_{\mathrm{h}}=-\frac{d p _{r}}{dr},\; \mathscr{F}_{\mathrm{g}}=-\frac{\nu(r)'}{2}(\rho+p_{r}),\; \mathscr{F}_{\mathrm{e}}=E(r)e^{\frac{\lambda (r)}{2}}\sigma(r).
\end{equation*}
where $\mathscr{F}_{\mathrm{a}}$, $\mathscr{F}_{\mathrm{h}}$, $\mathscr{F}_{\mathrm{g}}$, and $\mathscr{F}_{\mathrm{e}}$ denote the anisotropic force, hydrostatic force, gravitational force, and electric force respectively. The balancing development of the anisotropic, hydrostatic, gravitational, and electric forces can be perceived from the Fig. \textbf{10}. The total effect of the $\mathscr{F}_{\mathrm{a}}$, $\mathscr{F}_{\mathrm{h}}$, $\mathscr{F}_{\mathrm{g}}$, and $\mathscr{F}_{\mathrm{e}}$ demonstrates that our models are physically viable and stable.
\begin{figure}
\centering \epsfig{file=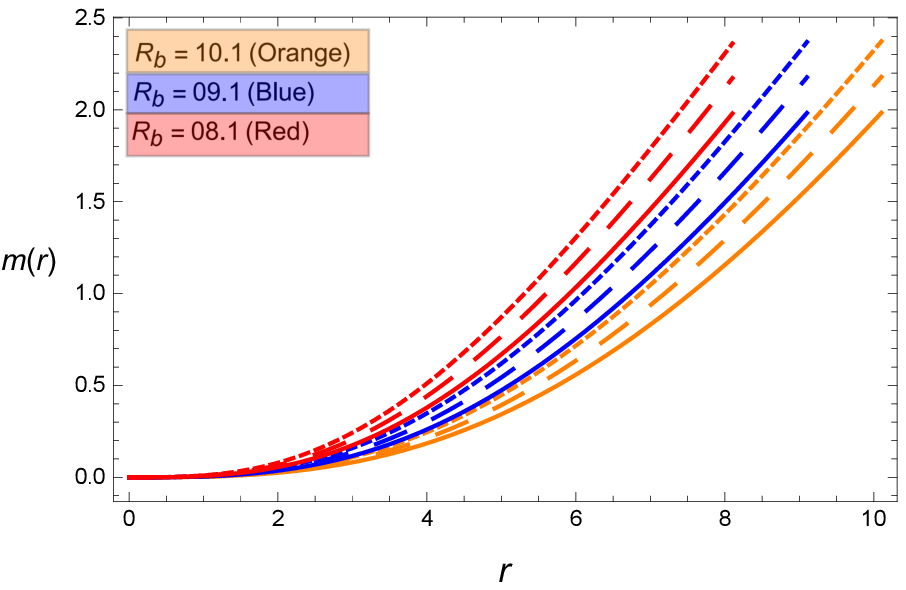, width=.48\linewidth,
height=2.3in}\epsfig{file=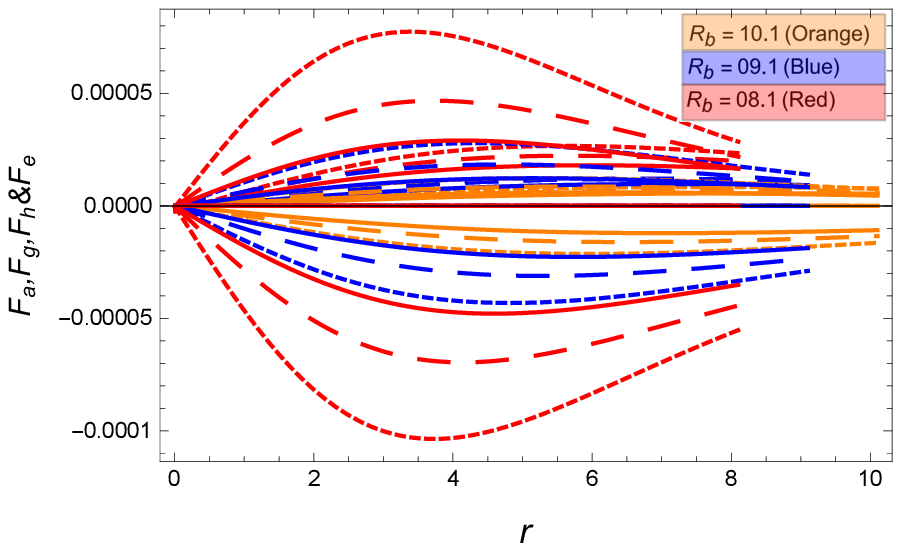, width=.48\linewidth,
height=2.3in} \caption{\label{Fig.10} Reflects the evolving profile of mass function and the TOV equation}
\end{figure}

\subsection{Causality stability analysis}

Now we discuss the speeds of sound for anisotropic pressure, i.e., radial, and tangential pressure components, represented by the $v^{2}_{r}$ and $v^{2}_{t}$, and are described as
\begin{equation}\label{38}
v_{r}= \sqrt{\frac{dp_{r}}{d\rho}}\;\;\;\;\;\;\Rightarrow  v^{2}_{r}= \frac{dp_{r}}{dr} \times \frac{dr}{d\rho},\;\;\;\;\;\;\;\;\;\;v_{t}= \sqrt{\frac{dp_{t}}{d\rho}} \;\;\;\;\;\;\Rightarrow  v^{2}_{t}= \frac{dp_{t}}{dr} \times \frac{dr}{d\rho}.
\end{equation}
It is supported by Fig. \textbf{11}, that the velocities $v^{2}_{r}$, and $v^{2}_{t}$ remain within the bound of $0\leq v_{r}$ \& $v_{t}\leq1$ as reflected in the plots of Fig. \textbf{12}. Moreover, the condition $-1\leq v^{2}_{t}-v^{2}_{r}\leq 0$ is also fulfilled \cite{43}.
\begin{figure}
\centering \epsfig{file=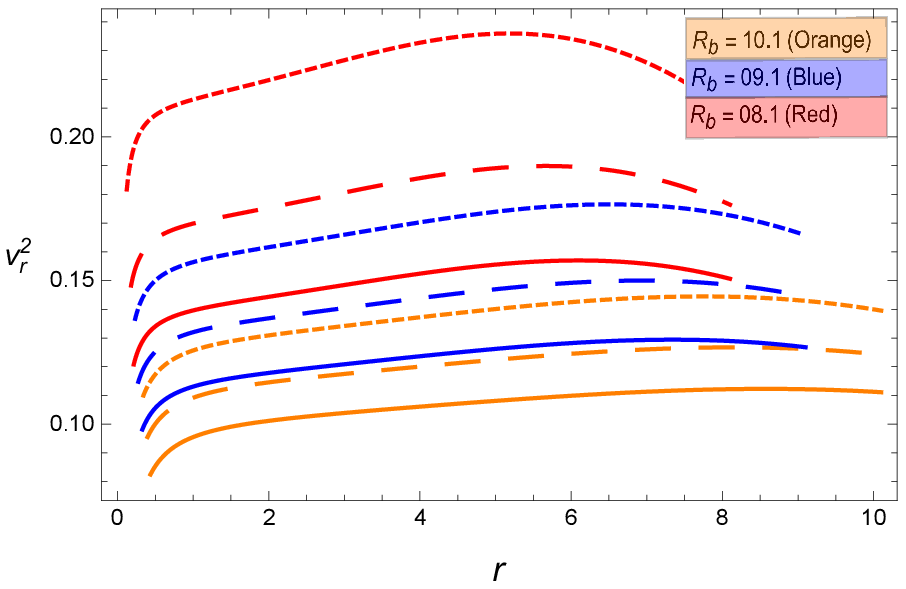, width=.48\linewidth,
height=2.3in}\epsfig{file=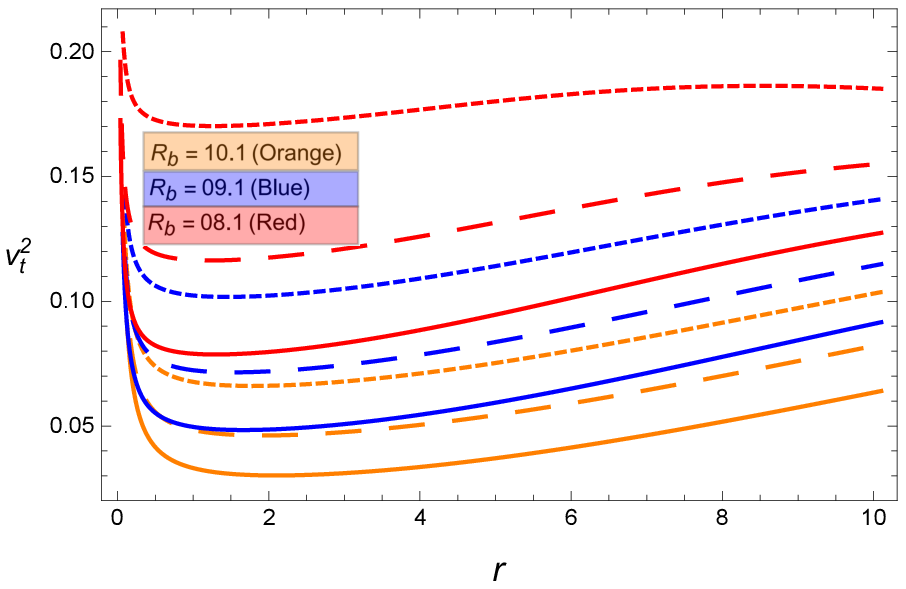, width=.48\linewidth,
height=2.3in} \caption{\label{Fig.11} Reflects the evolving profile of the behavior of speed of sound}
\end{figure}

\subsection{Adiabatic index stability analysis}

 Adiabatic index is represented here by $\Gamma_{r}$ and is defined as \cite{Hil}:
\begin{equation}\label{39}
\Gamma_{r}= \frac{\rho\times v^{2}_{r}}{p_{r}}\times \left(1+\frac{p_{r}}{\rho}\right)
\end{equation}
However, Chandrasekhar defined the stability condition for perfect fluid with reference to the following inequality \cite{Chand,Chand1}
\begin{equation}\nonumber
\Gamma\geq \frac{4}{3}+\frac{19}{42}\bigg(1-\big(\frac{1+{P}_0/\rho_0}{1+{P_0}^3/\rho_0}\big)^2\bigg).
\end{equation}
The modified version of this inequality for anisotropic fluid is given by
\begin{equation}\nonumber
\Gamma_{r}\geq \Gamma_{cr},
\end{equation}
where $\Gamma_{cr}= \frac{4}{3}+2\beta$ and
\begin{equation}\nonumber
\beta= \frac{19}{21}\bigg(1-\big(\frac{1+{P_r}_0/\rho_0}{1+{{P_r}_0}^3/\rho_0}\big)^2\bigg).
\end{equation}
It is worthwhile to mention here that Newtonian regime is recovered when $\beta=0$. It is interesting to notice that all our solutions obey the Chandrasekhar stability criteria as evident from Table-\textbf{II} and Fig. \textbf{12}.
\begin{center}
\begin{table}
\caption{\label{tab1}{Approximated values of $\rho_{0},\; p_{r_{0}}=p_{t_{0}},$ and $\Gamma_{cr}$.}}
\begin{tabular}{|c|c|c|c|c|c|c|c|c|}
    \hline
    \multicolumn{4}{|c|}{$R_{b}=10.10 (km)$} \\
    \hline
$M(M_{\odot})$ \;\;\;\;                  & $\rho_{0}$\;\;\;\                  &$p_{r_{0}}=p_{t_{0}}$\;\;\;               &$\Gamma_{cr}$\\
\hline
2.00\;\;\;\;                             &0.00075996           &0.0000455613      &1.42061 \\

2.20\;\;\;\;                             &0.000892062           &0.0000651053      &1.43519\\

2.40\;\;\;\;                             &0.00104262           &0.0000926685      &1.4514\\
    \hline
    \multicolumn{4}{|c|}{$R_{b}=09.10 (km)$} \\
    \hline
2.00\;\;\;\;                             &0.00111753          &0.0000839003      &1.43742\\

2.20\;\;\;\;                             &0.00132855           &0.000124249      &1.45584\\

2.40\;\;\;\;                             &0.00157511           &0.000184612      &1.47664\\
    \hline
    \multicolumn{4}{|c|}{$R_{b}=08.10 (km)$} \\
    \hline
2.00\;\;\;\;                             &0.00174857           &0.000173958     &1.46135\\

2.20\;\;\;\;                             &0.00211763           &0.000272728     &1.48582\\

2.40\;\;\;\;                             &0.00256392         &0.000435133     &1.51394\\
\hline
\end{tabular}
\end{table}
\end{center}

\begin{figure}
\centering \epsfig{file=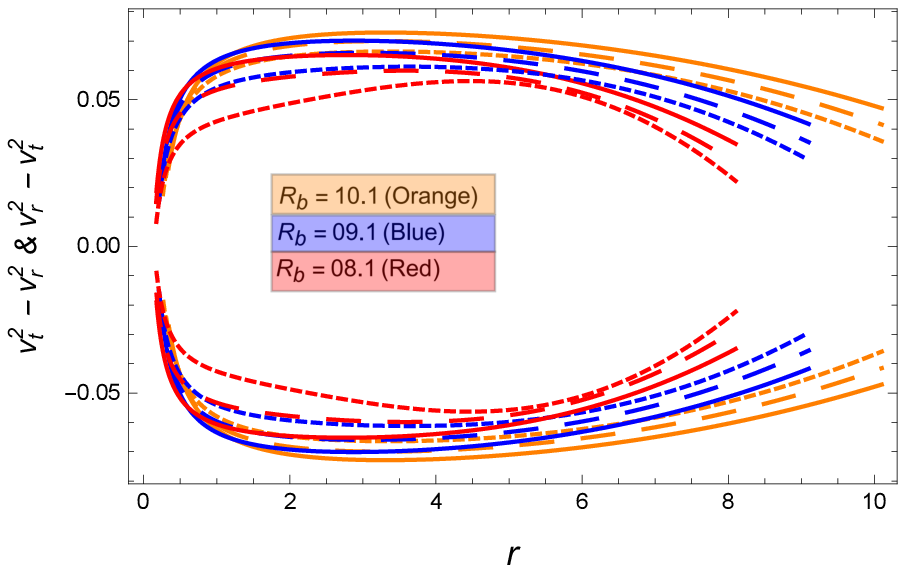, width=.48\linewidth,
height=2.3in}\epsfig{file=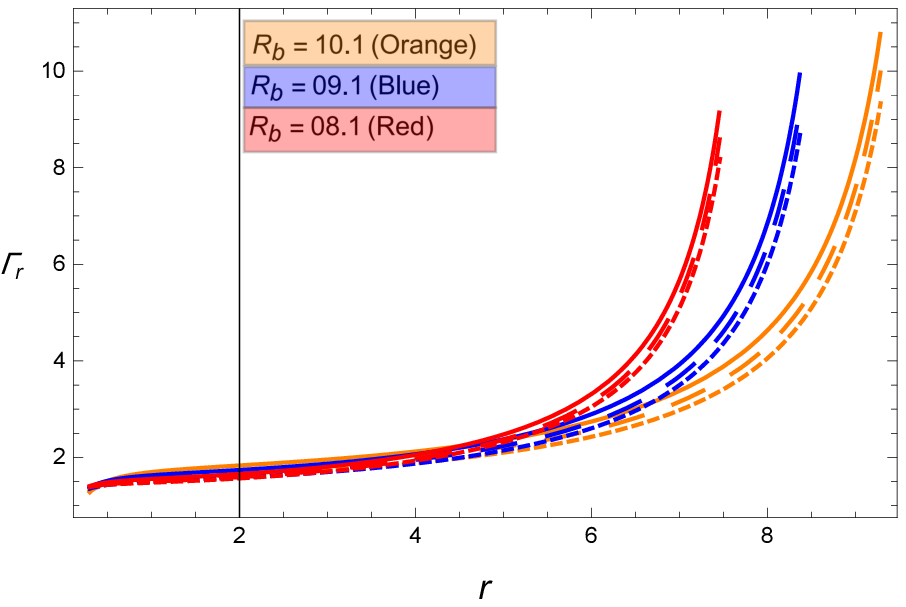, width=.48\linewidth,
height=2.3in} \caption{\label{Fig.12} Reflects the evolving profile of Abrea condition and adiabatic index}
\end{figure}

\section{Outlook}

In this study, we have explored compact stars solutions in background of Einstein-Maxwell's equations, with spherically symmetric geometry and anisotropic matter source. We have utilized the concept of Karmarkar condition by assuming the time metric co-efficient following the approach by Adler \cite{Adler}. Further, we have used the matching condition, by considering the Bardeen geometry as an outer spacetime \cite{Bardeen}. Bardeen geometry exhibits fascinating investigation. Implementation of this kind of approach turns out to be very interesting as this corresponds to the magnetic mono-pole gravitational remnants emerging from some particular non-linear electrodynamics. As far we know, this may be the pioneer work to introduce the Bardeen geometry in investigating the charged anisotropic spheres satisfying Karmarkar condition. A brief summary and prominent aspects of our study are exclusively pointed out as:
\begin{itemize}
\item
For the present analysis, we choose three different ranges of radial coordinate $r$, i.e., $0<r\leq 8.10,\;0<r\leq 9.10 $ \& $ 0<r\leq 10.10$, which are colored by red, blue, and oranges lines respectively throughout this discussion. Further, we utilize three diverse values of $M$, i.e., $M=2.00,\;M=2.20 $ \& $ M=2.40$, which are mentioned by solid, dashed, and small dashed throughout this paper. The worked-out approximated values of the involved parameters can be seen in the table-\textbf{I} for all the cases. \\
\item
 The crucially important physical parameters, i.e., $\rho$,$\;p_{r},\;p_{t},\; \triangle,\; \sigma,\; $$d\rho/dr, \;dp_{t}/dr$ and $dp_{r}/dr $. have plotted and analysed in detail with reference to their role in the emergence of compact structures under the Bardeen geometry.
Stability of the found  solutions has been discussed. Energy constraints, i.e.,  $DEC,\;$$SEC,\;$$WEC$, and $NEC$,, equation of state parameter, i.e., $w_r$ and $w_t$, generalized TOV equation via different forces, i.e., $\mathscr{F}_{\mathrm{a}}$, $\mathscr{F}_{\mathrm{h}}$, $\mathscr{F}_{\mathrm{e}}$ and $\mathscr{F}_{\mathrm{g}}$, causality condition, i.e., $0\leq v_{r}~\&~ v_{t}\leq1$, and adiabatic index, i.e., $\Gamma_{r}$ have also been graphed and analysed in detail. \\
\item
From  Fig. \textbf{1} it is perceived that $g_{rr}(r =0)=1$ and $g_{tt}(r =0)\neq0$, which suggests that the Karmarkar condition is physically acceptable to investigate the stability of charged compact stars. The Fig. \textbf{1} reflects the information of $\rho$ for three different ranges of $r$. It acquires throughout the positive profile for all values of $M$.\\
\item
As far the pressure components are concerned, it is revealed from Fig. \textbf{2} that $p_r$ bahaves positive for $R>r$, while the $p_t$ shows positive evolution throughout the configuration of stars. From Fig. \textbf{3}, it may be noticed that the electric field becomes zero at the center and then increases with positive nature outward while the charge density is decreasing from center to boundary.\\
\item
The anisotropy parameter $\triangle$ exhibits positive evolution in this study. The growing anisotropy function hints the presence of compact structures. as depicted in Fig. \textbf{4}. The gradients $d\rho/dr, \;dp_{t}/dr$ and $d p_{r}/dr$ are seen negative, i.e., $\frac{d\rho}{dr}< 0,\;\frac{dp_{r}}{dr}< 0,\;\frac{dp_{t}}{dr}< 0$ for $R\geq r$ and $\frac{d\rho}{dr}= 0,\;\frac{dp_{r}}{dr}= 0,\;\frac{dp_{t}}{dr}= 0$ for $r=0$ in the current analysis.\\
\item
The energy conditions $SEC$, $WEC$, $NEC$, and $DEC$, i.e., $\rho$, $p_r$, $p_t$, $\rho-p_{r}$, $\rho-p_{t}$, $\rho-p_{r}-2p_{t}$, are seen justified in this work. Further, their graphic behavior can be revealed from Fig. \textbf{6} and Fig. \textbf{7}. The parameters $w_r$ and $w_t$ remain positive inside the compact stars and fulfill the constraints $0\leq w_r~\&~ w_t\leq1$.\\
\item
The red-shift function $Z(r)$ in our case satisfies the criteria given by Bohmer and Harko \cite{40} and Ivanov \cite{41}. In our current analysis, $Z(r)\leq 3.100$, validating the physical acceptability of our study. $u(r)$, the compactness parameter stays within the Buchdahl limit \cite{42}. Also, the plots of the mass function $m(r)$ has acquired the increasing positive profile.\\
\item
The balancing behavior of the $\mathscr{F}_{\mathrm{a}}$, $\mathscr{F}_{\mathrm{h}}$, $\mathscr{F}_{\mathrm{e}}$, and $\mathscr{F}_{\mathrm{g}}$ forces can be noticed from the Fig. \textbf{10}. The speeds of the sounds $v^{2}_{r}$ and $v^{2}_{t}$ for compact stars, both satisfy the condition $0\leq v_{r}~\&~v_{t}\leq1$. Further, the Abrea condition \cite{43}, $-1\leq v^{2}_{t}-v^{2}_{r}\leq 0$ is also attained and can be seen from Figs. \textbf{11-12}. The profile of adiabatic index $\Gamma_{r}$ as evident from Fig. \textbf{12} that $\Gamma_{r}$ has shown monotonically increasing trend and greater than $\frac{4}{3}$ in this current study.\\\\
\end{itemize}


\end{document}